\documentclass[twocolumn]{aastex61}

\usepackage{amsmath,amstext}
\usepackage[figure,figure*]{hypcap}
\usepackage{newtxmath} 
\usepackage[utf8]{inputenc}
\usepackage[T1]{fontenc}
\usepackage{verbatim}

\received{}
\revised{}
\accepted{}
\submitjournal{ApJ}

\shorttitle{SFADI technique}
\shortauthors{Li Causi et al.}

\begin{document}

\title{SFADI: the Speckle-Free Angular Differential Imaging method.}

\correspondingauthor{Gianluca Li Causi}
\email{gianluca.licausi@oa-roma.inaf.it	}

\author{Gianluca Li Causi}
\affiliation{INAF Istituto di Astrofisica e Planetologia Spaziali - Via Fosso del Cavaliere 100, Roma, Italy}
\affiliation{INAF Osservatorio Astronomico di Roma - Via Frascati 33, 00078, Monte Porzio Catone (RM), Italy}
\affiliation{ADONI Adaptive Optics National Lab of Italy}

\author{Marco Stangalini}
\affiliation{INAF Osservatorio Astronomico di Roma - Via Frascati 33, 00078, Monte Porzio Catone (RM), Italy}
\affiliation{ADONI Adaptive Optics National Lab of Italy}

\author{Simone Antoniucci}
\affiliation{INAF Osservatorio Astronomico di Roma - Via Frascati 33, 00078, Monte Porzio Catone (RM), Italy}
\affiliation{ADONI Adaptive Optics National Lab of Italy}

\author{Fernando Pedichini}
\affiliation{INAF Osservatorio Astronomico di Roma - Via Frascati 33, 00078, Monte Porzio Catone (RM), Italy}
\affiliation{ADONI Adaptive Optics National Lab of Italy}

\author{Massimiliano Mattioli}
\affiliation{INAF Osservatorio Astronomico di Roma - Via Frascati 33, 00078, Monte Porzio Catone (RM), Italy}
\affiliation{ADONI Adaptive Optics National Lab of Italy}

\author{Vincenzo Testa}
\affiliation{INAF Osservatorio Astronomico di Roma - Via Frascati 33, 00078, Monte Porzio Catone (RM), Italy}
\affiliation{ADONI Adaptive Optics National Lab of Italy}



\begin{abstract}
We present a new processing technique aimed at significantly improving the angular differential imaging method (ADI) in the context of high-contrast imaging of faint objects nearby bright stars in observations obtained with extreme adaptive optics (EXAO) systems.
This technique, named "SFADI" for "Speckle-Free ADI", allows to improve the achievable contrast by means of speckles identification and suppression. This is possible in very high cadence data, which freeze the atmospheric evolution. Here we present simulations in which synthetic planets are injected into a real millisecond frame rate sequence, acquired at the LBT telescope at visible wavelength, and show that this technique can deliver low and uniform background, allowing unambiguous detection of $10^{-5}$ contrast planets, from $100$ to $300$ mas separations, under poor and highly variable seeing conditions ($0.8$ to $1.5$ arcsec FWHM) and in only $20$ min of acquisition. A comparison with a standard ADI approach shows that the contrast limit is improved by a factor of $5$. We extensively discuss the SFADI dependence on the various parameters like speckle identification threshold, frame integration time, and number of frames, as well as its ability to provide high-contrast imaging for extended sources, and also to work with fast acquisitions.
\end{abstract}


\section{Introduction} \label{sec:intro}
Direct imaging of exoplanets \citep[see for example][to mention a few]{2014IAUS..299...46M, 2015ESS.....320305B, 2015ESS.....320304C, 2015ApJ...800L..24H} have recently become possible thanks to both technological advances of extreme adaptive optics (EXAO) systems \citep[see for instance][]{cavarroc2006fundamental, 2006Msngr.125...29B, 2008SPIE.7015E..18M, esposito2010first, 2012SPIE.8447E..0BK, davies2012adaptive, 2014SPIE.9148E..03B, 2014PNAS..11112661M, 2014IAUS..299...32C, 2016arXiv160905147P}, and the introduction of high efficiency post-processing techniques like angular differential imaging \citep[ADI;][]{2006ApJ...641..556M, 2007ApJ...660..770L}, Karhunen-Lo{\`e}ve image decomposition \citep[KL;][]{2012ApJ...755L..28S}, or principal component analysis \citep[PCA;][]{2012MNRAS.427..948A}.
These methods allow the post-facto reduction of the stray light from the central star on the surrounding areas, with the aim of increasing the contrast and to reveal nearby sources, at angular separations of the order of a few hundreds of milli-arcsec. Indeed, the combination of EXAO and post-facto techniques yields impressive results in terms of faint source detectability \citep{2014ApJ...786...32M, 2016arXiv160905147P}.
While EXAO is by far the most important ingredient to boost the image contrast, ADI and other post-facto techniques are fundamental to get the most out of the data. The main idea behind them relies on the subtraction of the point spread function (PSF) of the central object in order to increase the contrast in the surrounding region. This is done by estimating the PSF itself. This task is usually accomplished by considering the median PSF, estimated throughout an observation composed of a temporal sequence of images, and by subtracting it from each frame of the sequence. It is worth noting that this PSF estimate may suffer from the intrinsic rapid variations of the seeing conditions. In other words, the estimated PSF may not be the ideal model at each instant, especially in the presence of highly varying seeing conditions, and this may result in suboptimal results.

The adoption of very fast acquisition systems, providing exposures of one or a few milliseconds, can freeze the atmospheric turbulence evolution, and offer a solution to this shortcoming. Very recently, \cite{2016arXiv160905147P} and \cite{doi:10.1117/1.JATIS.3.2.025001} have highlighted the importance of fast cadence in reducing residual jitter and in the application of post-facto techniques, by exploiting new $1$ ms cadence data acquired by the SHARK-VIS Forerunner experiment at LBT.

Here we present a new post-facto contrast enhancing method optimized for very high cadence imagers like the SHARK-VIS Forerunner. This technique is based on the identification and removal of residual speckles in milliseconds exposure images, and can be seen as a "speckle-free" extension of the standard ADI technique. For such reason, we refer to this technique as to "Speckle-Free ADI" (SFADI).\\

\section{Data set}
The data set used in this work consists of a $20$ minutes acquisition of $1.2\times10^6$ sequential $1$ ms exposure images of the target Gliese 777 (see left panel of Fig. \ref{Fig:SFADI_Concept}), acquired with the SHARK-VIS Forerunner experiment at LBT on June 4, 2015. The SHARK-VIS Forerunner experiment consisted in a set of short test observations performed at the LBT telescope to verify its EXAO system performance at visible wavelengths ($600 < \lambda < 700~$ nm) between February and June 2015 \citep{2016arXiv160905147P}. The experimental setup is minimal and composed by only two optical elements before the detector: one divergent lens to get a super sampling (twice the Nyquist limit) of the PSF and a $40$ nm FWHM filter centered at $630$ nm. The AO control and wavefront sensing is left to the LBTI Adaptive Optics subsystem (FLAO) fed through a $50\%$ beam splitter. The pixel scale is set at $3.73$ mas, and the imager is a Zyla sCMOS camera manufactured by Andor Inc\footnote{http://www.andor.com/}. During the acquisition, the LBTI-AO system \citep{esposito2010first} was correcting 500 modes in closed loop. The AO frequency was $990$ Hz with a loop delay of $3$ ms. In this conditions the closed loop $0$ db bandwidth is $59$ Hz. The seeing FWHM was in the range $0.8$ to $1.5$ arcsec, and no field de-rotator is employed on the mount to correct for the sky rotation, in order to implement an ADI-like approach to the data reduction. For more information about the data set and the acquisition system we refer the reader to \cite{2016arXiv160905147P} and \cite{doi:10.1117/1.JATIS.3.2.025001}.\\

\section{The standar ADI and the new SFADI method} \label{sec:SFADI}

The so-called ADI technique \citep{2006ApJ...641..556M} is based on a three-step concept: i) all the frames of the acquisition sequence are co-registered ((through FFT phase correlation in our case), ii) the median of all the frames is computed, and used as the PSF model that is subtracted from each single frame of the sequence, and iii) the PSF-subtracted residuals are de-rotated to compensate for the field rotation, and median-combined to obtain the resulting image, mostly free from the contribution of the bright star PSF.

If the PSF were not varying among frames, the ADI would lead to a complete subtraction of the starlight, because the median PSF across the frames would be a good representation of the instant PSF at all times. In the practice, however, as shown in \cite{2016arXiv160905147P}, the seeing evolution can be highly variable throughout the observation, so that the median of the frames does not match the instantaneous PSF of any frame, producing strong residuals in the ADI result, which limits the achievable image contrast.



The SFADI method overcomes some of the above limitations through the full exploitation of the fast ($1$ ms) cadence of the data.

\begin{figure*}
  \centering
  \includegraphics[trim=0cm 0cm 0cm 0cm, clip, height=6.cm]{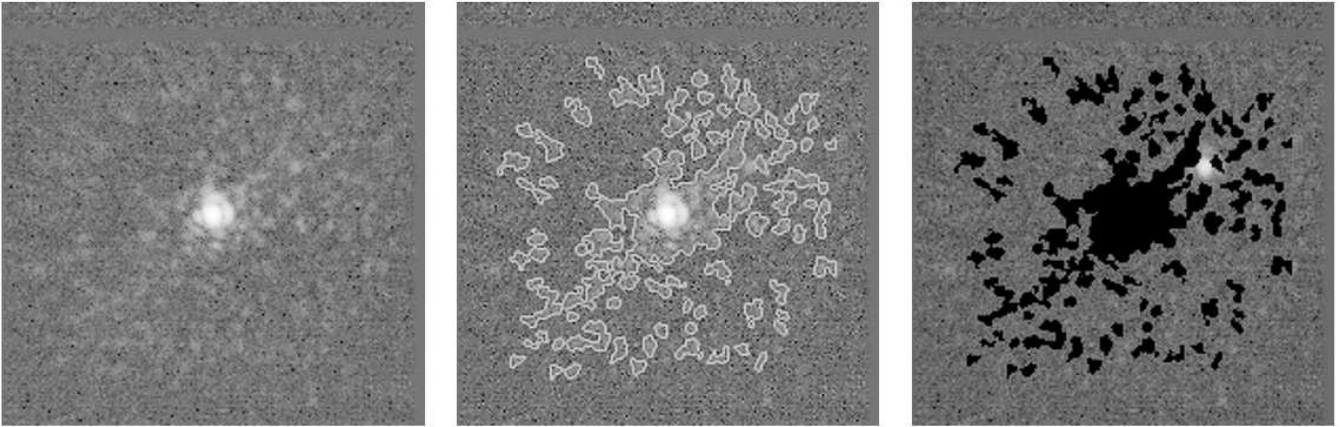}
   \caption{SFADI concept: the speckles in each original frame of the acquisition sequence (left) are identified by the SWAMIS feature recognition code (center), and masked out to leave only the starlight-unaffected background regions (right), where possible sources (we injected here a bright $10^{-1}$ contrast planet for much clarity) can be imaged with no light pollution from central star.}
    \label{Fig:SFADI_Concept}
   \end{figure*}

In fact, the frame by frame PSF evolution is due to the rapid variation of the adaptive optics residuals, which appears as a pattern of speckles changing shape with a timescale of less than $10$ ms, at visible wavelengths, as shown in \cite{doi:10.1117/1.JATIS.3.2.025001}. Hence, if the frame rate is fast enough, i.e. of the order of a few milliseconds, all the speckles are basically frozen (see Sec. \ref{sec:tracking}) and they appear as compact sharp features in each frame, not smoothed by their fast movement (left panel in Fig. \ref{Fig:SFADI_Concept}). In this case a suitable image recognition algorithm can be employed for identifying and suppressing them in each frame. As the star light is in practice completely contained in the speckles, the background pixels in between them will ideally contain no star light at all. Fig. \ref{Fig:SFADI_Concept} qualitatively illustrates this concept. Thus, if we exclusively use these background pixels in a standard ADI processing, we expect to have no speckle-induced artifacts in the final result, as shown in second panel of Fig. \ref{fig:ADI_SFADI_Full_Dataset}, because we only combine speckle-free regions from each frame.\\

\subsection{Speckle identification} \label{sec:tracking}

The first step of our method is the identification of the speckles in each single frame. This step is accomplished by using the SWAMIS code \citep{2007ApJ...666..576D}. This code was originally written for the identification and tracking of small-scale magnetic elements in the solar photosphere \citep{2008ApJ...674..520L, 2010ApJ...720.1405L, 2013ApJ...774..127L}, a task conceptually similar to that of the identification of faint speckles in AO data \citep{2014A&A...561L...6S, 2015A&A...577A..17S}. In short, the code uses a double-threshold clump identification scheme that allows to label small scale features. While the higher threshold is not necessary for our purpose and can be set to an arbitrary high value, the second lower threshold allows to isolate clusters of pixels above the noise, and is typically set at $2-3 \sigma$, where $\sigma$ is the standard deviation of the noise. The $\sigma$ of the noise distribution of our dataset, computed in a region $300$ mas away from the central source, is $2.3$ ADU.

In addition to this threshold, another stringent constraint is used to reject noise. Indeed, only clumps with a size larger than the resolution angle are selected (i.e. not smaller than the average FWHM of $\sim4$ pixels of the PSF core in our dataset). This implies that only structures with a spatial scale of the order of the PSF core are identified, thus ruling out the possibility to include noise features and also allowing identification of very faint speckles next to noise level.


This code was already used by \cite{stangalini2017} on exactly the same data to study the statistics of speckles in AO images at visible wavelengths. The main conclusions of this analysis were that $90\%$ of the AO residual speckles have a lifetime shorter than $5$ ms. This means that, at least at visible wavelengths, a very fast cadence of the order of $1$ KHz is required for the succesfull application of the SFADI technique or any other deconvolution method \citep[e.g.][]{2011OExpr..19.1975J} that relies on the freezing of the atmospheric turbulence evolution.

For further details on the application of the SWAMIS code for the identification of AO residual speckles we refer the reader to \cite{stangalini2017}.

\begin{figure*}
  \centering
  \includegraphics[trim=0cm 0cm 0cm 0cm, clip, height=5.5cm]{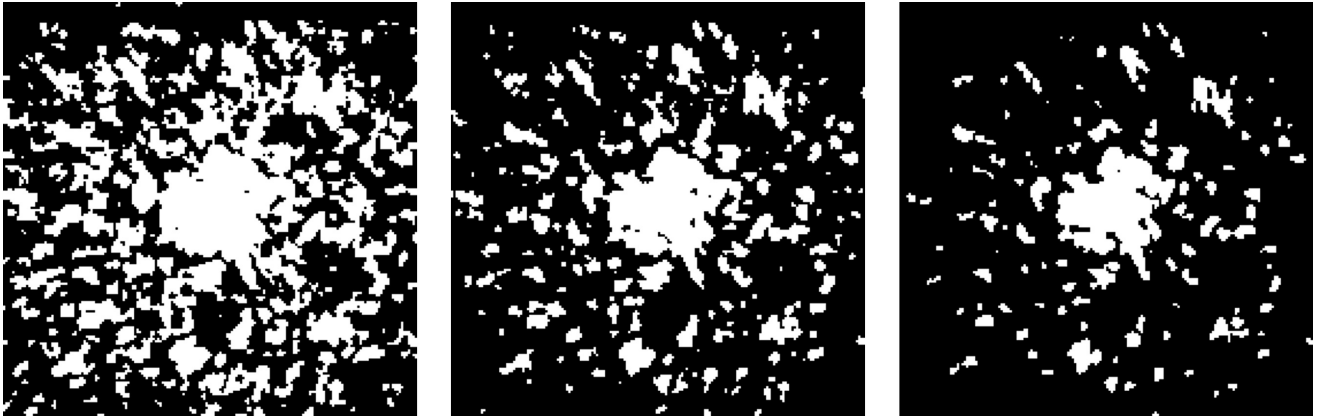}
   \caption{Speckle identification obtained by the SWAMIS code with a threshold of $2$, $3$ and $4$ ADU (from left to right, respectively).}
    \label{Fig:SWAMIS_Thresholds}
   \end{figure*}

The  output of the SWAMIS code are binary masks that identify clusters of pixels belonging to the same speckle. In Fig. \ref{Fig:SWAMIS_Thresholds} we show examples of  such identification for different thresholds (i.e. $2$, $3$, and $4$ ADU): the lower the intensity threshold the larger the number of speckles identified outside the central core of the PSF. In contrast, larger thresholds allow to isolate different speckles in the PSF core, which otherwise would appear as a single extended feature.\\

\subsection{Frames combination} \label{sec:frame_combining}

After identification of speckles in all frames, we compute the median background across the whole sequence, like one usually does in the ADI, but excluding the masked pixels. Then we take this median background as representative of the instant inter-speckle background at all times, and subtract it from each frame, still excluding the pixels belonging to identified speckles. Finally we perform the numerical de-rotation of these background-subtracted and masked frames, thus obtaining the stack of sky-aligned frames to be combined in the resulting image.

\begin{figure*}
  \centering
  \includegraphics[trim=0cm 0cm 0cm 0cm, clip, height=5.4cm]{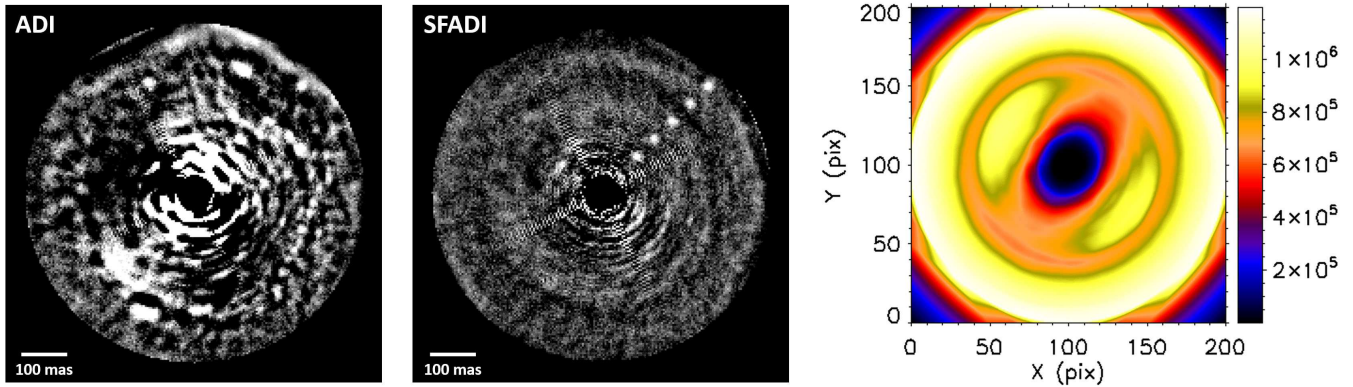}
   \caption{ADI and SFADI results, for the full $1.2\times10^6$ frames sequence described in the text, in which five synthetic $2\times10^{-5}$ contrast planets at $100$ to $300$ mas separation were injected. ADI result is shown in left panel: only the planet at $300$ mas is visible, but impossible to distinguish if previously unknown, because its flux is comparable to the residual features in the background. SFADI result is shown in central panel using the same grayscale: all the planets are clearly distinguishable over a much lower and uniform residual background. Regions closer than $35$ mas and outside $340$ mas, which contain the strongest artifacts, have been masked for clarity. The right panel depicts the map of the number of combined frames per pixel in the SFADI combination.}
    \label{fig:ADI_SFADI_Full_Dataset}
   \end{figure*}

In the standard ADI this combination of the aligned frames must be unavoidably computed through a median operator, because outlying pixel values are present wherever the median PSF differs from the instant PSF. This usually leads to artifacts in the combined result as shown in left panel of Fig. \ref{fig:ADI_SFADI_Full_Dataset}.

\begin{figure}
\includegraphics[width=8cm]{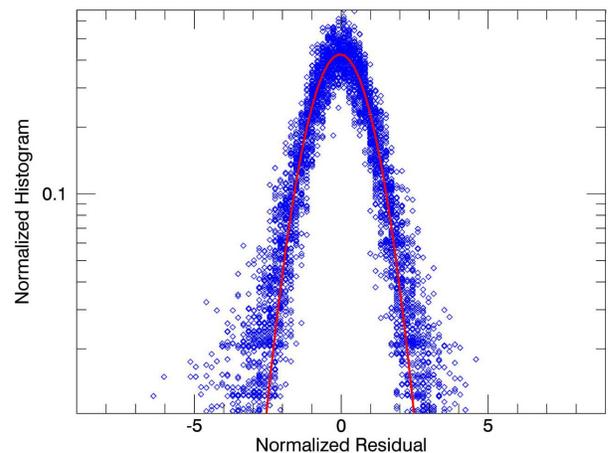}
\caption{Gaussian distribution of pixel values in the sky-aligned frames of the SFADI, for $100$ samples at $150$ mas distance from the star. Gaussian fit is overplotted in red.}
\label{fig:ADI_SFADI_Stat}
\end{figure}

In contrast, the distribution of pixel values in the sky-aligned frames of the SFADI is Gaussian (Fig \ref{fig:ADI_SFADI_Stat}) because no outliers are present in the inter-speckle background. Such situation allows us to combine the sky-aligned frames of the SFADI by means of a simple arithmetic average, which delivers a very uniform and artifact-free background in the resulting image, as we show in the central panel of Fig. \ref{fig:ADI_SFADI_Full_Dataset}.

This also yields a significant gain in computation time, as we adopt a straightforward running average, only slightly modified for excluding the masked pixels, which is much faster and less memory consuming than the median operator.

Fig. \ref{fig:ADI_SFADI_Full_Dataset} shows a direct comparison of the ADI and the SFADI result for our full $1.2\times10^6$ frames sequence, in which we injected five planets $2\times10^{-5}$ times fainter than the star, with separations of $50$ mas from $100$ to $300$ mas off-axis, and aligned at $45$ deg, i.e. along the direction where the ADI residual is found to be worse. In the same figure, we also show the map of the number of combined frames per pixel (right panel), as in the SFADI a different number of sky-aligned frames are effectively combined for each pixel of the final image.

For this figure we injected planets fainter than the ADI detection limit of $5\times10^{-5}$ achieved by \cite{2016arXiv160905147P}, using the same data. The improvement over the standard ADI method is evident. This will be quantitatively characterized in the following section.\\


\subsection{SFADI characterization} \label{sec:SFADI_tests}

In this section we present a quantitative comparison of the SFADI performances with respect to a standard ADI and to the photon limit, for different number of frames and different frame integration times.

The metric that we adopted for performing this comparison relies on the ratio of the aperture photometry at known planet positions to the standard deviation of same aperture photometry at each pixel location in the background.

Both the ADI and SFADI combined images have been processed, before photometry, by subtracting a $3$ FWHM box-median filtering and a $360$ degrees angular median, in order to flatten the large scale background variations.

\begin{figure*}
  \centering
  \includegraphics[trim=0cm 0cm 0cm 0cm, clip, width=16.cm]{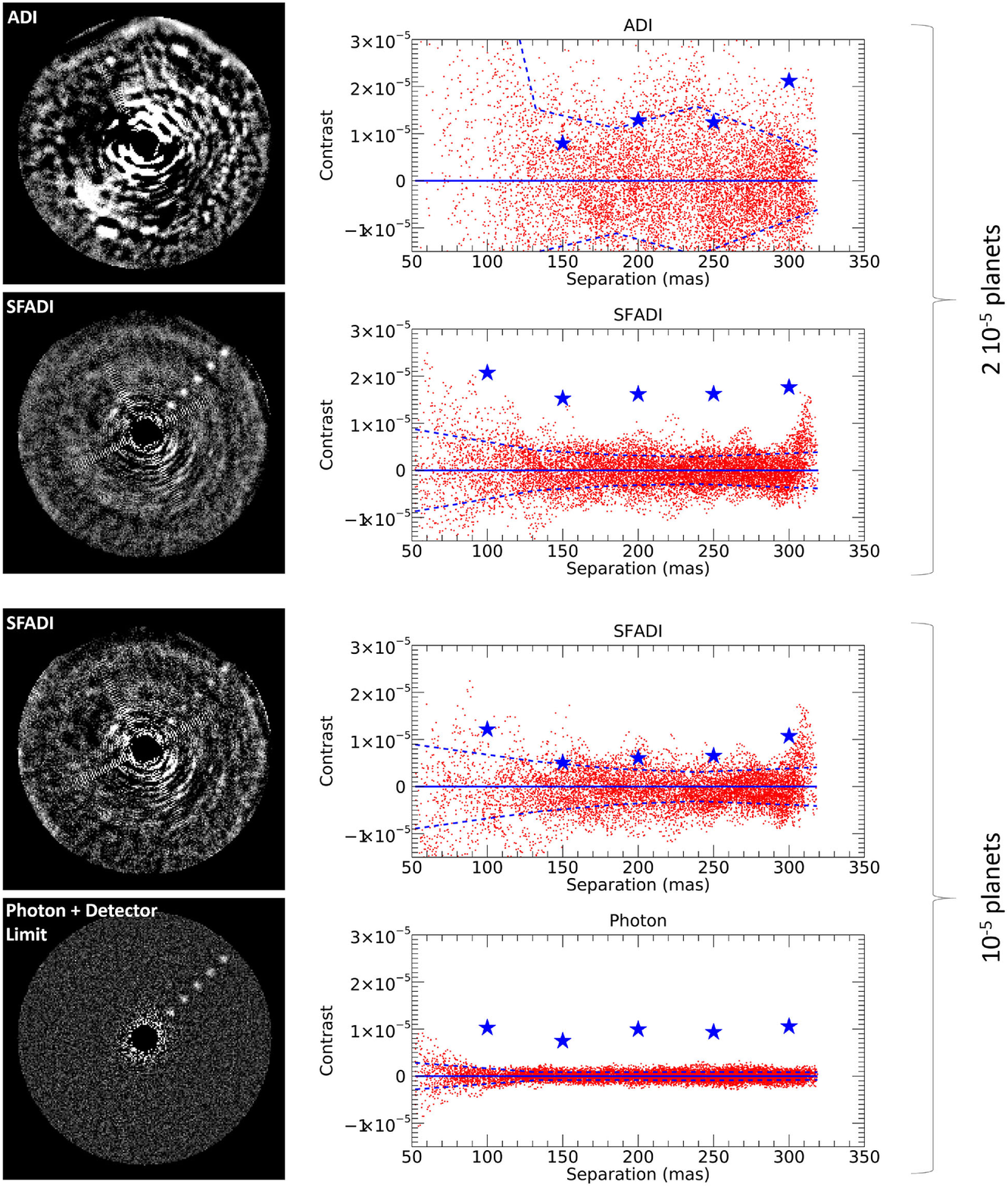}
   \caption{Planet detection performance for $2\times10^{-5}$ contrast planets with the ADI (top) and SFADI (second row): planets aperture photometry (blue star symbols) are compared to background aperture photometry at each pixel location (red dots), whose standard deviation is indicated by the blue dashed lines as a function of separation from central star. Third row: same analysis for $10^{-5}$ contrast planets. The theoretical photon limit, with $10^{-5}$ contrast planets, is shown in the bottom row, for comparison.}
    \label{fig:ADI_SFADI_Detection}
   \end{figure*}

Top and second row in Fig. \ref{fig:ADI_SFADI_Detection} show the planets photometry (stars) compared with the background photometry (red dots). It is shown that in the ADI the $2\times10^{-5}$ planets lie at only $1\sigma$ above the background, while in the SFADI they lie at $3\sigma$ above background. The third row of same figure shows that even a planet contrast as low as $1\times10^{-5}$ is still well detectable at all distances with the SFADI method. It is worth noting that an advanced detection algorithm would recognize planets even fainter than these, as it would also recognize the point sources against the background residuals, which have a very different shape.

Finally, in the bottom row of Fig. \ref{fig:ADI_SFADI_Detection} we show the same analysis applied to a numerical simulation, which takes into account the actual number of aligned frames per pixel reported in Fig. \ref{fig:ADI_SFADI_Full_Dataset}, and in which the star PSF is subtracted perfectly, i.e. only leaving detector and photon noise after subtraction. This condition represents the theoretical detection limit, and clearly demonstrates how the SFADI method is able to reach a residual level as low as $3$ times the photon limit.\\

\subsubsection{Dependence on frame exposure}

Such high performance of the SFADI method, which relies on the very short exposure time ($1$ ms) of the single frames of our dataset, would not be possible with longer integration times, as the speckle pattern has a typical lifetime of the order of a few ms \citep{stangalini2017}, as we mentioned in Sec. \ref{sec:tracking}.

\begin{figure*}
  \centering
  \includegraphics[trim=0cm 0cm 0cm 0cm, clip, width=16.cm]{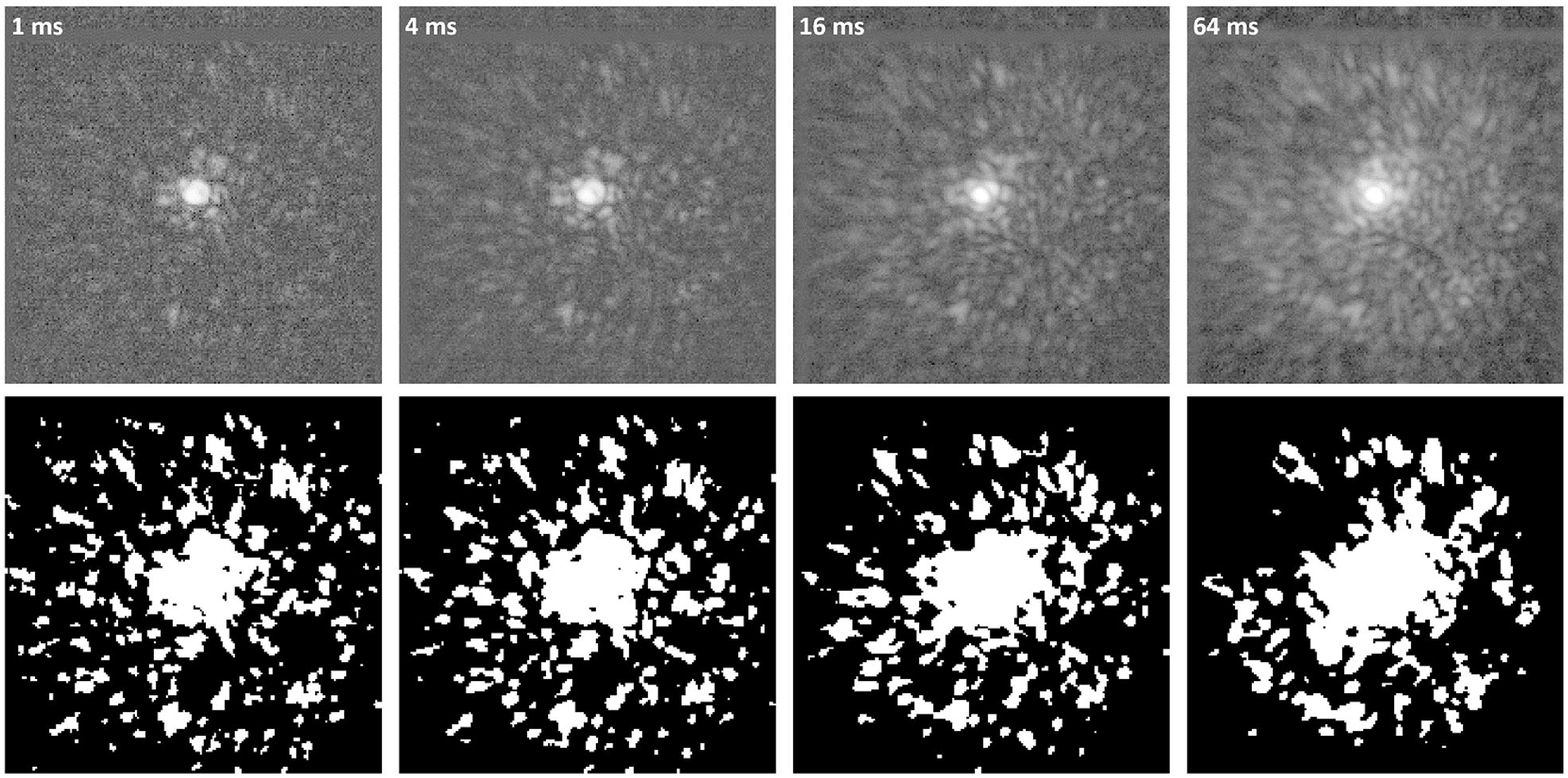}
   \caption{Speckles (top) and speckle recognition masks (bottom) for frame exposures of $1$, $4$, $16$, and $64$ ms (from left to right respectively).}
    \label{fig:SFADI_Frame_Rate}
   \end{figure*}

In order to better illustrate the SFADI performance with respect to the frame exposure time, we simulated longer integration times by summing up together $4$, $16$, and $64$ sequential frames and applying the SWAMIS identification code to them, as displayed in Fig. \ref{fig:SFADI_Frame_Rate}. As the frame integration increases, the quality of speckle recognition decreases. This is because multiple speckles are merged together.

\begin{figure*}
  \centering
  \includegraphics[trim=0cm 0cm 0cm 0cm, clip, width=14.5cm]{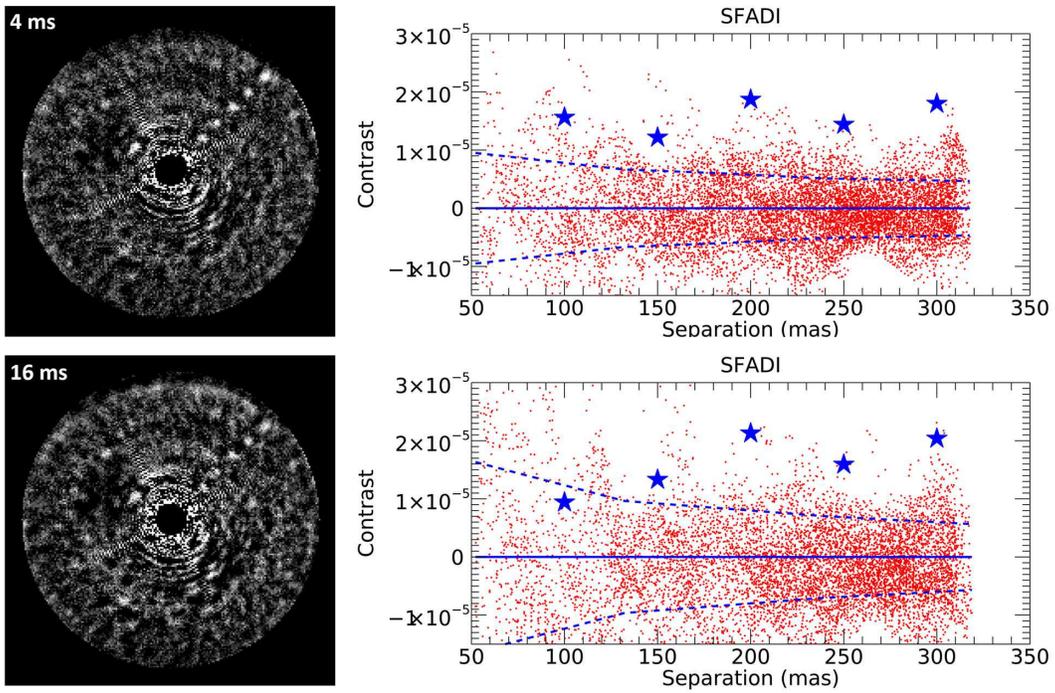}
   \caption{SFADI results for frame exposures of $4$ ms and $16$ ms (top and bottom respectively), to be compared with the $1$ ms case in second panel of Fig. \ref{fig:ADI_SFADI_Detection}.}
    \label{fig:SFADI_Binned_4_16}
   \end{figure*}

Fig. \ref{fig:SFADI_Binned_4_16} quantitatively shows the increment in background artifacts for the $4$ ms and $16$ ms cases and how the $2\times10^{-5}$ planets becomes undetectable, with respect to the $1$ ms case in second panel of Fig. \ref{fig:ADI_SFADI_Detection}. This demonstrates that a KHz frame rate is mandatory for boosting the SFADI to its maximum performance.\\

\subsubsection{Dependence on the number of frames}

  \begin{figure*}
  \centering
  \includegraphics[trim=0cm 0cm 0cm 0cm, clip, width=14.cm]{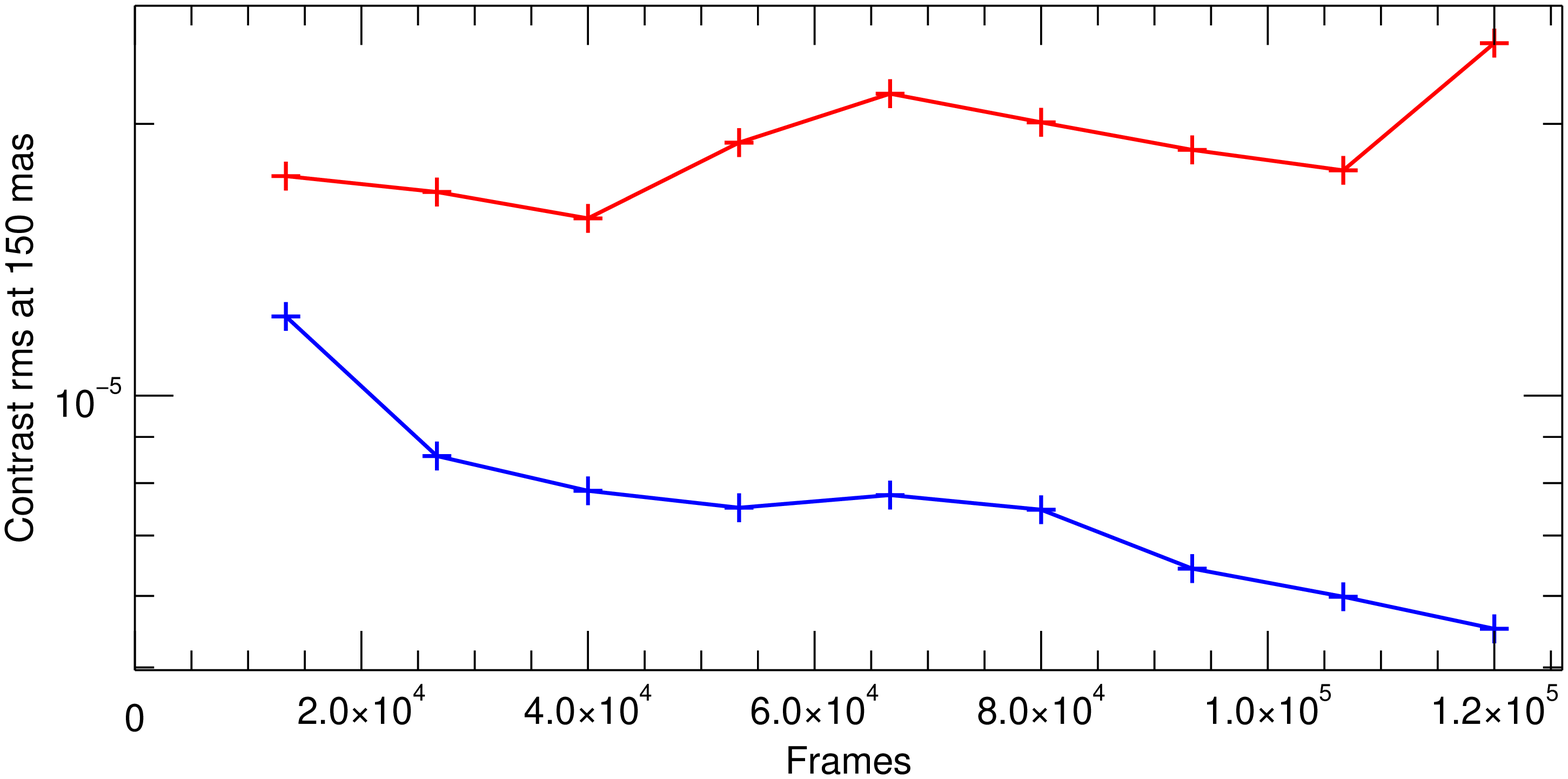}
   \caption{ADI (red) and SFADI (blue) standard deviation of background photometry as a function of the number of frames of the acquisition sequence. Note: a subset of $1$ frame every tenth has been used here for speeding up the computation time.}
    \label{fig:ADI_SFADI_Frame_Number}
   \end{figure*}

We finally compared the SFADI and ADI behavior as a function of the number of images in the frames sequence, in order to clarify whether acquiring a longer sequence effectively increases the quality of the result.

This is not obvious, because in the presence of highly variable seeing conditions, like in our dataset, the difference between the estimated PSF and the instantaneous PSF increases with the number of frames.

For this reason, even if the detector and photon noise percentage contribution decreases by increasing the acquisition time, the residuals of the standard ADI do not decrease. In fact, in the case of our dataset the seeing regime was different for different intervals of the observation, so that the ADI residuals increase as the image sequence gets longer (see Fig. \ref{fig:ADI_SFADI_Frame_Number}).

On the contrary, the SFADI method is not sensitive to the seeing variation, because the instantaneous PSF is canceled out by speckle masking in each frame. Consequently, the standard deviation of background residuals keeps reducing with time as depicted by the blue line in Fig. \ref{fig:ADI_SFADI_Frame_Number}.\\

\subsection{Extended sources and fast acquisitions: from SFADI to SFI} \label{sec:SFADI_extended_sources}

The speckle suppression technique that we described is not only useful for detecting point sources like extra-solar planets, binary stars, or background stellar objects, but it also opens the possibility to perform high-contrast imaging on extended sources, such as circumstellar disks and jets or AGN structures, and around multiple bright stars.

The main obstacle for the observation of extended features with standard ADI is that, if the size of the feature is not significantly smaller than the arc subtended by field rotation at the feature's position, the median PSF model contains an angularly-smoothed ghost of the feature itself, thus leading to self-subtraction in the final residual.
The same also happens for planets when the frame sequence is not long enough to locate the planet in the last frames at a sufficiently different position angle from the first frames.

On the contrary, if the speckles are masked, as we do in the SFADI, the remaining background virtually contains no starlight, but only light from the astronomical source of interest, so that we can skip the stage of median PSF subtraction and directly combine the de-rotated masked frames, revealing the extended feature in the background. In this case we skip any angular differential operations, so that we simply call this procedure "SFI", for "Speckle-Free Imaging".

\begin{figure*}
  \centering
  \includegraphics[trim=0cm 0cm 0cm 0cm, clip, width=17.cm]{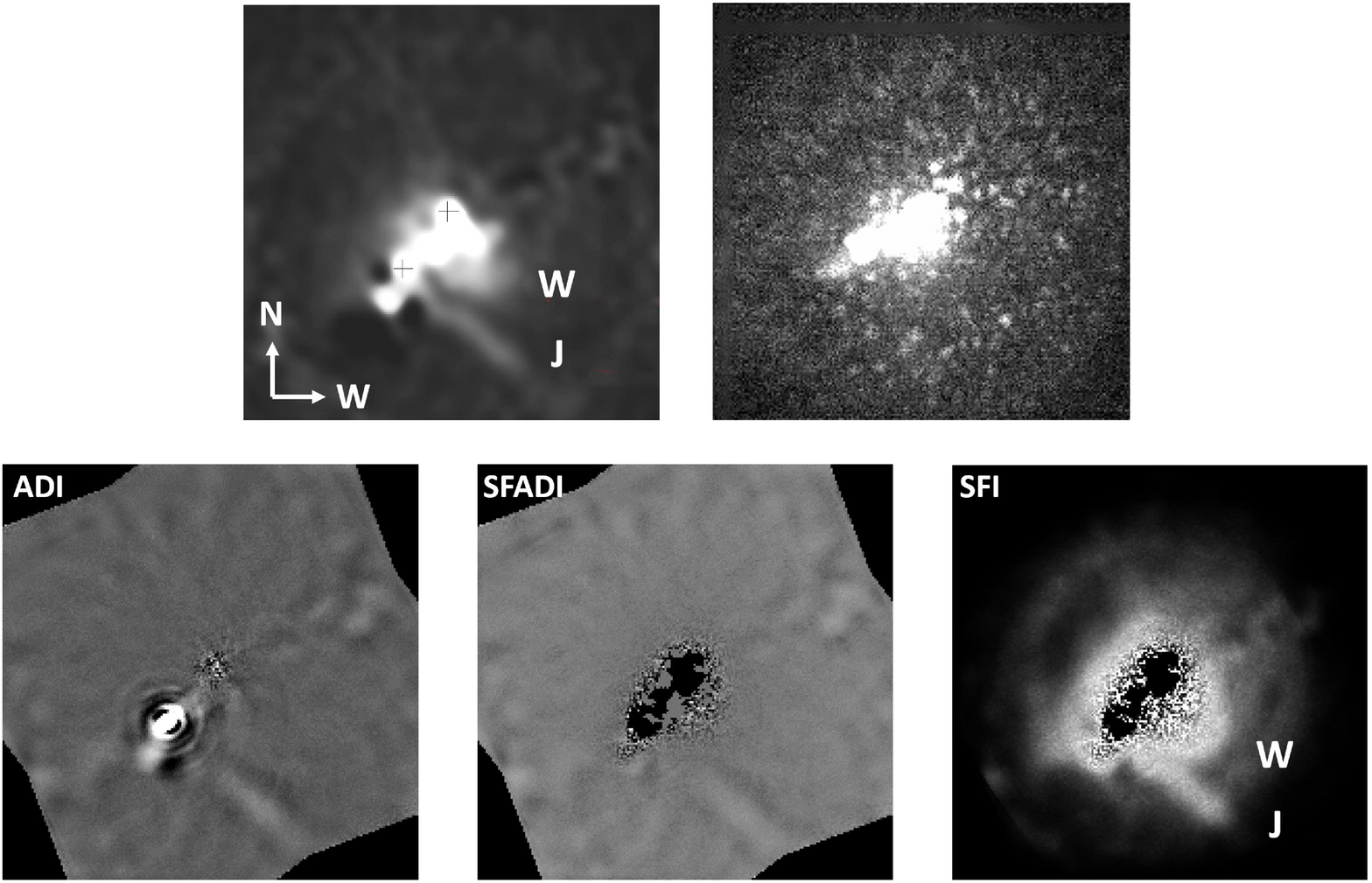}
   \caption{The extended circumstellar structure ($10^{-4}$ contrast) around the ZCMa binary (whose flux-scaled model, based on \cite{2016A&A...593L..13A}, is shown in top-left panel) as viewed after convolution with the PSF of a single frame of the SHARK-VIS Forerunner sequence (top-right). In the bottom panels the results of the application of ADI, SFADI, and SFI methods are shown (from left to right, respectively). The W and J labels indicate the wide wind from the primary NW component, and the collimated jet from the secondary star, respectively.}
    \label{fig:ADI_SFI_Extended_Source}
   \end{figure*}

As an example of this procedure we simulated the observation of a binary star surrounded by extended circumstellar structures, which we report in Fig. \ref{fig:ADI_SFI_Extended_Source}. The image model that we have considered (first panel in Fig. \ref{fig:ADI_SFI_Extended_Source}) was built from the deconvolved image of the young binary Z CMa obtained by \cite{2016A&A...593L..13A} using $[OI]\lambda6300$ line SPHERE/ZIMPOL observations, scaled to same flux of the SHARK-VIS Forerunner target star. In the circumstellar structure we can identify a wide-angle stellar wind emitted by the NW component (labelled with W) and a collimated jet driven by SW component (labelled with J), whose flux are scaled to an average constrast of about $10^{-4}$ with respect to the brightest NW component.

Both components of this binary produces speckles in each frame of the sequence (second panel in Fig. \ref{fig:ADI_SFI_Extended_Source}) and both the speckle systems are identified by the SWAMIS speckle mask. The result, as the lower panels clearly show, is that both the W and J extended structures are well imaged by the SFI technique (last panel), while the ADI and the SFADI contain strong artifacts by self-subtracted image components.\\

\begin{figure*}
  \centering
  \includegraphics[trim=0cm 0cm 0cm 0cm, clip, width=16.cm]{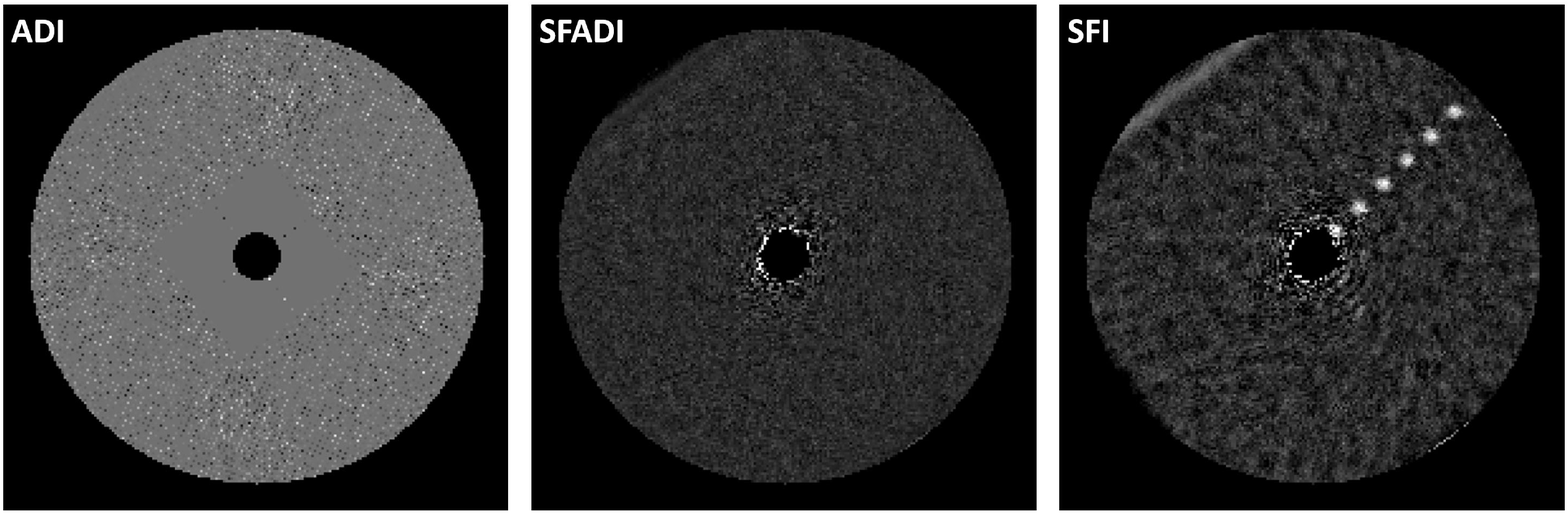}
   \caption{Simulation of six $5\times10^{-4}$ contrast planets, from $50$ to $300$ mas separation, with $5$ seconds acquisition after ADI, SFADI, and SFI post-processing methods (from left to right respectively). Note the visibility of the inner planet at only $50$ mas of separation from the central star.}
    \label{fig:ADI_SFI_Planet_Fast}
   \end{figure*}

Finally, Fig. \ref{fig:ADI_SFI_Planet_Fast} depicts the case of a very short acquisition sequence lasting $5$ seconds only, which does not provide enough field rotation for disentangling the simulated $5\times10^{-4}$ contrast planets, which in fact are not recovered by either ADI (left) and SFADI (center), but which are well detected by the SFI method (right). This means that the SFI technique can also be used to have a real-time preview of the SFADI result just after the very first frames of a long acquisition sequence.\\

\section{Conclusions} \label{sec:Conclusions}

We introduced the new Speckle-Free Angular Differential Imaging (SFADI) technique for high contrast imaging, based on speckle identification and masking on millisecond cadence acquisitions.

The presented method substantially improves the standard ADI performances for both planet detection and extended sources nearby single-conjugated adaptive optics guide stars.

Applying SFADI to a real $20$ minutes sequence of $1$ ms exposure frames acquired at LBT, for which the standard ADI contrast limit is $5\times10^{-5}$ at $100$ mas, we reach a contrast limit of $1\times10^{-5}$ despite the poor and highly variable seeing conditions ($0.8$ to $1.5$ arcsec) of the observation.

We also presented extended tests for characterizing how the SFADI results depend on the frame integration time and on the number of acquired frames. We show that, thanks to the fast KHz frame rate, the SFADI performances approach the theorethical photon limit, and that they continue improving as the acquisition time increases, suggesting that very high contrast limits can be reached with long observations.

At last, we introduce the Speckle-Free Imaging (SFI) technique, which basing on the SFADI concept opens the high contrast imaging to extended objects and to fast acquisitions.\\

\acknowledgments
This study has been supported by the ADONI Italian National Laboratory for Adaptive Optics and developed to fully exploit the performances of the SHARK-VIS high-contrast imager for LBT, currently under construction.

\vspace{5mm}
\facilities{LBT}
\software{SWAMIS}

\bibliographystyle{yahapj}
\bibliography{lib.bib}

\end{document}